\documentstyle[prb,aps,preprint,epsf]{revtex}

\def\limfunc#1{\mathop{\rm #1}}%
\begin{document}
\begin{center}
{\bf {\large \ SMECTIC-LIKE PHASE FOR MODULATED XY SPINS IN TWO DIMENSIONS}}
\\{\today} \vspace{.5 cm}

  {\it  M. Gabay}\\
\vspace{0.5cm}

Laboratoire de Physique des Solides
Laboratoire associ\'e au CNRS, Universit\'e de Paris-Sud,\\B\^atiment 510,
91405 Orsay Cedex, France\\
\vspace{0.5cm}                                                      

{\it M. Benakli}\\ 
\vspace{.5cm}            
                                                                                
Department of Physics, Condensed Matter
Section, ICTP, P.O. Box 586, 34014 Trieste, Italy\\                             
   \vspace{0.5cm}                                                               
 
  {\it  and W.M. Saslow}\\
\vspace{0.5cm}

Department of Physics,
Texas A\&M University,
College Station, Texas 77843-4242\\
\end{center}
\maketitle
%%%%%\maketitle <--- this places the abstract on a separate page%%%%%%%%%%%%%%

\begin{abstract}

The row model for frustrated \textsc{XY} spins on a triangular lattice in 2D is
used to study incommensurate (\textsc{IC}) spiral and commensurate (\textsc{C})
antiferromagnetic (AF) phases, in the regime where a
(\textsc{C})-(\textsc{IC}) transition occurs. 
Using fluctuating
boundary conditions and specific histogram techniques, a detailed Monte
Carlo (MC) study reveals more structure in the phase diagram than found
in previous MC simulations of the full parameter space. On the (C) side,
equilibrium configurations consist of alternating stripes of spiral phases of   
opposite chirality separated by walls of the (\textsc{C}) phase. 
For this same parameter regime, thermodynamic quantities are computed  
analytically using the NSCHA, a generalization of the self consistent
harmonic approximation appropriate for chiral systems. On the
commensurate side of the (C)-(IC) boundary, NSCHA predicts an
instability of the (C) phase. This suggests that the state is spatially
inhomogeneous, consistent with the present MC result: it resembles the
smectic-A phase of liquid crystals, and its existence  
implies that the Lifshitz point is at ${T=0}$ for modulated XY\
spins in 2D.  The connection between frustrated XY systems and the vortex state
of strong type II superconductors suggests that the smectic phase may 
correspond to a vortex liquid phase of superconducting layers.  

{\bf PACS Number 75.10.Hk}

 \end{abstract}

\newpage

\section{Introduction}

Frustration is an ubiquitous phenomenon in condensed matter physics. It
occurs whenever several ground states of a system compete at different 
length scales. Examples of such a situation are non-interacting electrons
in a tight binding potential subjected to a uniform magnetic field\cite{DRH},
networks of superconducting wires\cite{HKM} or of Josephson
junctions\cite{IJHS} in a field and spins with competing
interactions\cite{HTD}. In particular, frustrated magnetic systems have been
used in the quantum case as realizations of the spin liquid state\cite{KL}
(advocated in the context of high $T_c$ superconductors).  In the classical
case, XY (${O(2)}$) spins model the vortex state of layered, strong type II
superconductors\cite{LS,BL}. Frustration manifests itself by the existence of
chiral variables. The effect of this additional (${Z}_2$) symmetry on phase
transitions is still an open debate.  
For fully frustrated models  there remains to establish whether the the ${Z}_2$
and the ${O(2)}$ symmetries are broken at different temperatures or at the same
temperature{\cite{FOOT1,BENGRA,BZG,JRS,O3}}. In the context of $2D$ helimagnets
this issue comes up when one studies the commensurate-incommensurate 
((C)-(IC)) transition\cite{HTD}: on the (C) side, the state is non chiral
 whereas on the (IC) side chirality is coupled to the XY variables.
Insight into this particular problem can be gained
by studying the phase diagram of the row model, an anisotropic frustrated 2D XY
model on a triangular lattice:
it is a generalization of the fully frustrated XY\ model on the
triangular lattice (FFTXY) where all the bonds strengths J are multiplied by
$\eta $ in the horizontal direction\cite{ZSG,ZSGB,KAWA}  (the FFTXY\ model 
corresponds to $\eta =1$).

In order to study this system\cite{SaslowCLF} and other (IC) 
structures\cite{CS}, a
MC\ algorithm with "self determined (Fluctuating) Boundary Conditions'' (FBC)\
was developed.  The resulting $\eta$ versus ${T}$ phase diagram indicated a 
continuous (C)-(IC) transition line starting at $\eta =0.5$ for ${T=0}$ and
ending at a Lifshitz point\cite{HLS} (LP) for $\eta _L\simeq 0.62$ and
$T_L\simeq 0.42J$.  
Besides, increasing ${T}$ at fixed $\eta$ 
($0.5<\eta <\eta _L$) produced the following sequence of phases: an (IC) state
at low ${T}$; then, across the (C)-(IC) transition, at ${T}_{C-IC}$, one moves
into the (C) phase; lastly one reaches the paramagnetic (P) boundary at ${T}_P$.  In this
process one of the eigenvalues of the spinwave stiffness matrix decreases
uniformly as ${T}$ varies from zero to ${T}_{C-IC}$, vanishes at ${T}_{C-IC}$,
increases again in the (C) phase and becomes zero above ${T}_P$. 

These MC results raise an issue, because they yield  
a thermodynamically stable commensurate state for 
${T}\gtrsim {T}_{C-IC}$ and also because they predict a finite temperature LP:
from the standpoint of critical phenomena, the (C) phase is in the
 same universality
 class as the 
ferromagnetic XY model; in the vicinity of the LP 
one may analytically compute the bare (unrenormalized)
stiffness constant and one finds that it is very small. In this regime, the KT
renormalization group equations\cite{KT,PM}  show that 
vortex-antivortex pairs are unbound, implying that the (C) phase is
thermodynamically unstable near the (C)-(IC)\ transition; so 
renormalization
group predicts 
a reentrant (P) phase  
 and thus a zero temperature Lifshitz point,
at variance with the Monte Carlo results. Another indication that the LP may
occur at $T=0$
comes from the study of $2D$ modulated ${O(N)}$ spin systems exhibiting
 a (C)-(IC) transition;
renormalization group analysis
predicts\cite{ADDJ} that the LP is at ${T=0}$ whenever $N>2$; 
numerical studies show that this also holds if $N=1$ (ANNNI model)\cite{WS}.
Interpolating to the case $N=2$ one might have then expected a zero temperature
 LP for $2D$ XY systems: using a
phase-only Hamiltonian, Garel and Doniach indeed reached this conclusion for
the so-called $J_1-J_2$ model\cite{J1J2}.

The  present paper reconciles these a-priori conflicting results:

In section \ref{sec.MC}, we present a MC algorithm allowing to study
incommensurate and spatially inhomogeneous states\cite{boundaryhist}: 
it combines FBC and
specific boundary condition histograms  designed for FBC.  This approach allows us to
analyse the data near the (C)-(IC) transition.  Section \ref{sec.RM} presents 
a MC study of the row model for

\begin{figure}
\begin{center}
{\parbox[t]{10.5cm}{\epsfxsize 10.5cm
\epsffile{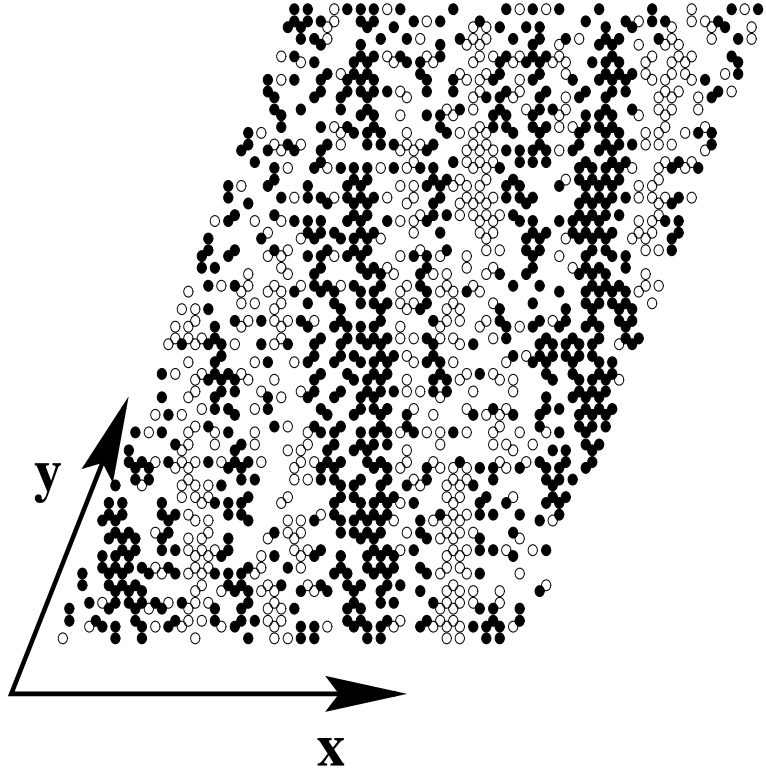}}
}
\end{center}
\protect\caption{
Snapshot of chiralities on each  plaquette of a $36^2$ triangular lattice.
$\eta=0.575$ and $T=0.4J$. Filled
circles represent plaquettes with the correct sign, i.e in the same
 chiral state
 as at $T=0$. Open circles correspond to plaquettes
 with the wrong sign, that is
 such that the chirality has changed compared to $T=0$. Plaquettes with zero 
chirality  (no symbol) are obtained in-between the two.
 One clearly sees a stripe 
structure of filled circles and open circles separated by domain walls of zero
chirality.
}
\label{VISURUB}
\end{figure}

\begin{figure}
\begin{center}
{\parbox[t]{11.5cm}{\epsfxsize 11.5cm
\epsffile{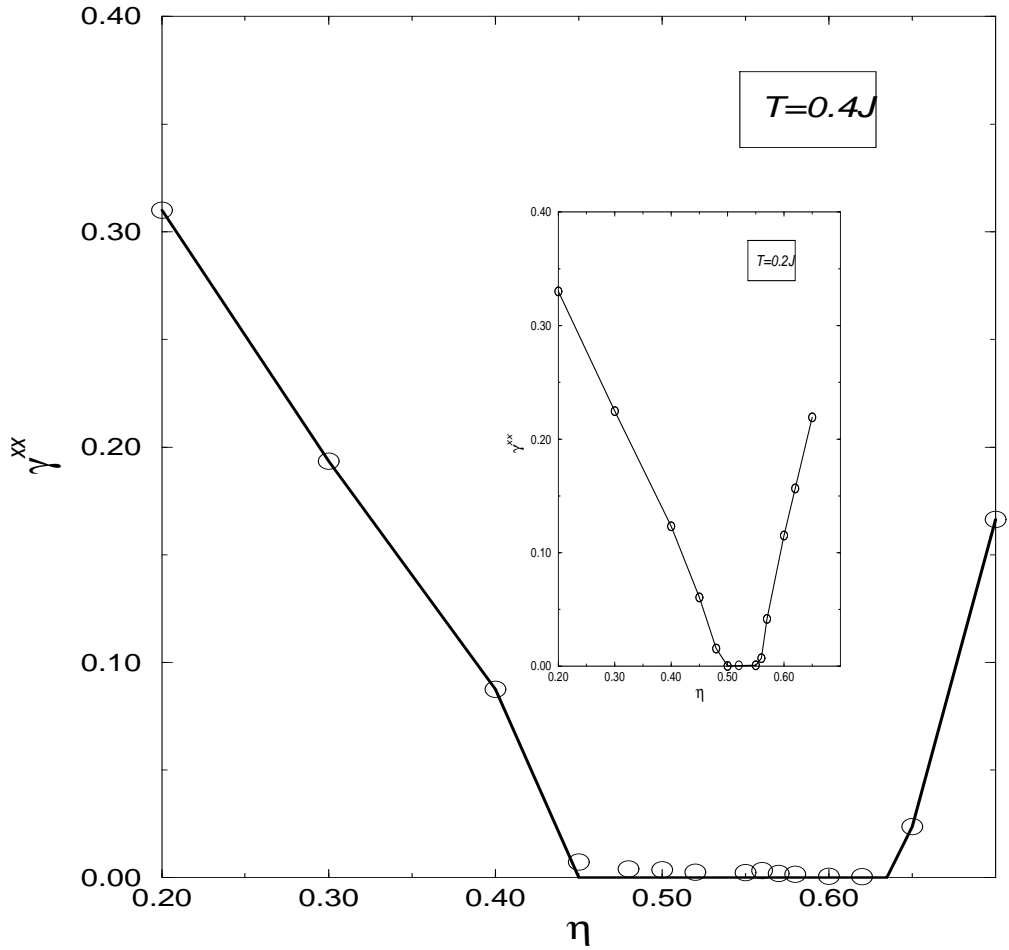}}
}
\end{center}
%\vspace{2cm}
\protect\caption{
MC data for $\gamma^{xx}$, versus $\eta$. The lattice size is $48^2
$.  $\gamma^{xx}$ is
obtained from the histogram in $\Delta$ modulo $2\pi\over L$. The region where
 $\gamma^{xx}=0$ corresponds to the domain of stability of the stripe phase.
$T$ is fixed: $T=0.4J$ (inset $T=0.2J$).
}
\label{GAMAETA}
\end{figure}

\begin{figure}
\begin{center}
{\parbox[t]{13.5cm}{\epsfxsize 13.5cm
\epsffile{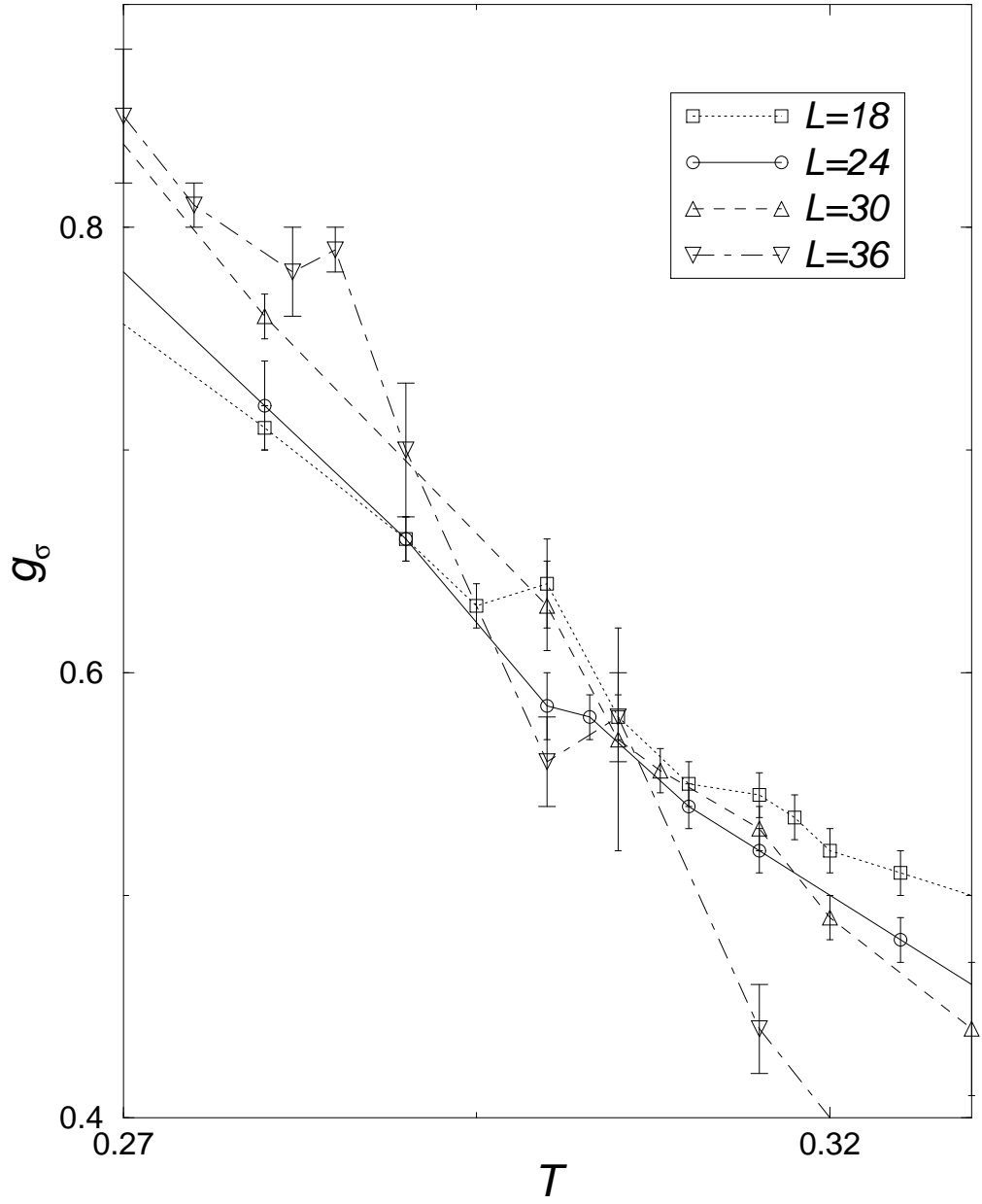}}
}
\end{center}
\protect\caption{
Binder order parameter $g_{\sigma}$ versus $T$ for various sizes (Eq.\ref{binder}).
}
\label{BINDEROP}
\end{figure}

\noindent
  $0.5<\eta <\eta _L$, for 
various ${T}$ and $\eta$ near the (C)-(IC) transition. Special attention
 is devoted to the ''(C) phase'' for ${T}\geq 
{T}_{C-IC}$. Our results suggest that the equilibrium structure is
 spatially inhomogeneous: Fig (\ref{VISURUB}) shows a striped structure,
corresponding to 
the coexistence of domains of opposite chirality separated by walls of the
collinear phase.  Such a state resembles the smectic-A
phase of liquid crystals. In this regime, we find that $\gamma^{xx}$ -- the
spin rigidity in the horizontal ($\eta$-bond) direction --
 is zero, whereas
$\gamma^{yy}$ -- the spin rigidity in the vertical direction -- is strictly
 positive
(Fig (\ref{GAMAETA})). 
  Stripes exist because  the coupling between phase and
chiral variables is relevant at all $T$ when $\eta>0.5$.  This coupling helps
explain why domains of the chiral phase are present for $T\geq {T}_{C-IC}$.
Moreover, fluctuations between a spatially homogenous state (the incommensurate
phase) and a spatially inhomogeneous spiral domain state (the striped phase) do
not allow simple scaling analysis of critical quantities at ${T}_{C-IC}$ (Fig
(\ref{BINDEROP})). 

These observations allow us to conclude that, in the phase diagram,
 the (C) and (IC) phases
are separated by a smectic-like phase, and only come in contact at $T=0$ and
$\eta=0.5$, so that the LP is indeed at ${T=0}$ for the 2D XY\ model.  On the
other hand there is no re-entrant (P) phase between the (C) and (IC) regions. 

Our numerical findings are further supported by analytic calculations,
presented in section \ref{sec.NSCHA}. These use the NSCHA method
 (New Self-consistent Harmonic Approximation), a recently developed variational
approach for frustrated systems\cite{BZG}.

\newpage

\section{Monte Carlo}\label{sec.MC}

\subsection{Fluctuating boundary conditions.}

For incommensurate
phases, the choice of periodic boundary conditions (PBC) in a MC simulation is not suitable,
since these break the magnetic symmetry of the system.

Instead, self-consistent boundary conditions, using FBC, have been proposed to
overcome the problem\cite{SaslowCLF,O1}. The main feature of  FBC is to add new
dynamical variables $\Delta_\alpha $ ($\alpha =1,2,...,D\;\;\;$ where $D$ is the dimensionality of the lattice )
corresponding to a shift at the boundaries. In equilibrium the new
''boundary variables'' $\Delta _\alpha $ will fluctuate around their
\textit{most probable value} $\Delta _\alpha ^0$.
For an $L\times L$ system of XY
spins on a lattice,
the FBC method amounts to imposing the following constraint on the phases $\theta(\vec r)$ 
of the spins, at the boundary 
  \begin{equation}
\theta (\vec r+nL\vec u_x+mL\vec u_y)=\theta (\vec r)+nL\Delta _x+mL\Delta _y
\label{phiclf}
\end{equation}
Using 
FBC allows us to preserve translational invariance: 
performing a change of variables 
\begin{equation}
\theta(\vec r)=\varphi (\vec r)+\vec \Delta .\vec r
\label{VARCHANG}
\end{equation}

 the constraint on $\varphi $ becomes
\begin{equation}\label{VARPERI}
\varphi (\vec r+nL\vec u_x+mL\vec u_y)=\varphi (\vec r)
\end{equation}
  In terms
of the new variable $\varphi $ the partition function of the $L\times L$
system with FBC is:

\begin{equation}
Z_{FBC}=L^2\int_{-\pi /L}^{\pi /L}d^2\Delta \left( \int ...\int_{-\pi }^\pi
\prod_id\varphi _ie^{-\beta .\left( -\frac 12\sum_{i,j}J_{ij}\cos (\varphi
_i-\varphi _j-\vec \Delta .(\vec r_i-\vec r_j))\right) }\right)  \label{ZCLF}
\end{equation}
$Z_{FBC}$ can be factorized as a product of a set of partition functions, $%
Z(\vec \Delta )$, each one corresponding to a fixed shift $\vec \Delta$ at
the boundaries:

\begin{equation}
Z_{FBC}=L^2\int_{-\pi /L}^{\pi /L}Z(\vec \Delta )d^2\Delta =L^2\int_{-\pi /L}^{\pi
/L}d^2\Delta e^{-\beta L^2f(\vec \Delta )}  \label{ZCLF2}
\end{equation}
where $f(\vec \Delta )$ is the $\frac{2\pi }L$ periodic free energy 
density associated with the shift  $\vec \Delta $ at the boundary: $f(\vec
\Delta )=-T\ln (Z(\vec \Delta ))/L^2$. 

For a system with a helical phase at low temperature,  $f(\vec\Delta )$ displays
a minimum for $\vec \Delta=\vec \Delta ^0$ and
 for a spiral phase, the pitch $\vec Q_0$, is the $\frac{2\pi }L$ determination of
 $\vec\Delta_0$ such that $\varphi (\vec r)\simeq 0$ in equilibrium (see Eq.\ref{VARCHANG}).
Since the main contribution to the integral (Eq.\ref{ZCLF2}) comes from
$\vec \Delta=\vec \Delta_0$,  the components $%
\gamma^{xx},\gamma^{yy}$ of the spin rigidity \cite{PM} are given by 

\begin{equation}
\gamma^{xx}=\rho \left. \frac{\delta ^2f(\vec \Delta )}{\delta \Delta
_x^2}\right| _{\vec \Delta ^0},\;\;\;\gamma^{yy}=\rho \left.
\frac{\delta ^2f(\vec \Delta )}{\delta \Delta _y^2}\right| _{\vec \Delta
^0} \label{gamma2}
\end{equation}
where $\rho $ is a (lattice dependent) geometrical factor.

At low $T$ and far from the (C)-(IC) boundary (where $\gamma^{xx}=0$), 
$\beta \gamma^{xx}>>1$ and $\beta\gamma^{yy}>>1$. 
Using Eq.\ref{ZCLF2} and Eq.\ref{gamma2}
 then gives\cite{SaslowCLF,boundaryhist}
 
\begin{equation}
\gamma^{xx}=\frac \rho {L^2\chi _{\Delta _x}},\;\;\;\gamma^{yy}=\frac
\rho {L^2\chi _{\Delta _y}}  \label{gammadelta}
\end{equation}
where $\chi _{\Delta _x}=\beta<(\Delta _x-\Delta^0 _x)^2>$ (resp. $\chi _{\Delta _y}=\beta<
(\Delta _y-\Delta^0 _y)^2>$ ) is the susceptibility
for $\Delta _x$ (resp. $\Delta _y$ ).

\subsection{Boundary condition histograms: $\Delta -$Histograms} \label{Del.hist}

In the previous section we showed that the partition function with FBC is 
a sum over partition functions $Z(\vec \Delta )$. A practical way to perform this sum is to count the number of
 configurations obtained for each of the allowed values of $\Delta _{x}$ and 
$\Delta _{y}$. Since $
\Delta _{x}$ and $\Delta _{y}$ are defined modulo $\frac{2\pi }L$ , this can be easily done by
histograms in $\Delta _{x}$ and $\Delta _{y}$, which we call
$\Delta -$histograms.

Denoting by $P(\vec\Delta)\equiv P(\Delta _x,\Delta _y)$ the probability distribution 
 for $\vec\Delta$,   
 the $\Delta $-histogram free energy density is obtained from: 
\begin{equation}\label{freenerhist}
f(\vec \Delta )=-\frac 1{\beta L^2}\ln \left( P(\vec \Delta) \right)
+Constant
\end{equation}
If $f(\vec \Delta )$ has a deep minimum for $\vec \Delta=\vec \Delta ^0$, the zeroes of the
 first derivative of the free energy yield the value of
$\vec \Delta ^0$ . The second derivatives of the free energy computed for
$\vec \Delta=\vec \Delta ^0$ give the components of the spinwave stiffness $ \gamma $,
by Eq.\ref{gamma2}. But even if $P(\vec\Delta)$ is not sharply peaked (see below),
histograms allow to compute
any thermodynamic observable as an average over $P(\vec\Delta)$. 

This algorithm is especially useful when {\it (i)} one approaches a critical
(C)-(IC) transition: $\vec \Delta$ undergoes large fluctuations and
 Eq.\ref{gammadelta} breaks down; histograms give much more accurate results
 and are well suited to scaling analysis {\it (ii)} equilibrium configurations
correspond to inhomogeneous structures: in that case, histograms yield multi-peak structures.
 For instance, if 
 domains of the (C) and (IC)
phases coexist near $T_{C-IC}$    
 the free energy will display  minima at $\vec\Delta =\vec0$ and at 
 $\pm \vec \Delta ^0$. 
\vskip 1cm

\section{Numerical analysis of the row model near the (C)-(IC)
 transition}\label{sec.RM}

Since the incommensurability is  only present in the $x$ ($\eta-$bonds) direction we used 
hybrid boundary conditions: PBC\ in the $y$ direction and FBC in the $x$
direction.  
A standard Metropolis algorithm was applied to
the spin angles and to the boundary shift in the $x$ direction.
 Lattices sizes ranged from $18^2$ to $48^2$ and the number of
 MCS/spin was of order $10^5 - 10^6$. Typically the first $10^4$ steps
 were discarded for equilibration. In contrast to 
our previous study of this system\cite{SaslowCLF}, $\Delta -$histograms
 were included here. These
were used to determine $Q_0$ (the $x$ component of the wavevector) as well as
  the spinwave stiffnesses along $x$ and $y$. In addition, we  monitored 

\begin{itemize}
\item
the staggered chiralities $\Sigma=\bigl<\sigma\bigr>$ with
\begin{equation}\label{sigmaMC}
\sigma =\frac 1{N_P}\sum_{\left\{ P\right\} }\frac{
\sum_{\left\langle kl\right\rangle \in P}\sigma _{kl}}{%
\sum_{\left\langle kl\right\rangle \in P}\sigma _{kl}(T=0)}
\end{equation}
 where $P$ refers to plaquettes in the same chiral state at $T=0$ 

\noindent
and
\begin{equation}\label{sigmaprimeMC}
\sigma _{kl}=\frac 1{2\pi}(\theta _k-\theta _l)
\end{equation}
(for Eq.\ref{sigmaprimeMC}, the angular determination of the term in parenthesis
 is taken in the interval $[-\pi,+\pi]$ (see Ref.\ref{LBZG})).

\item
the chiral susceptibility 
\begin{equation}\label{chiralxi}
\chi_{\sigma}=\frac 1{T}\Bigl<\sigma^2
 -\Sigma^2\Bigr>
\end{equation}

\item
the Binder order parameter for chiralities
\begin{equation}
g_{\sigma}=\frac 1{2}\Biggl[3-\Biggl({ \bigl<\sigma^4\bigr>\over{\bigl<\sigma^2\bigr>}}\Biggr)\Biggr]
\label{binder}
\end{equation}
\end{itemize}

\subsection{Study of the (C)-(IC) line at fixed $\eta$}

\begin{figure}
\begin{center}
{\parbox[t]{11.5cm}{\epsfxsize 11.5cm
\epsffile{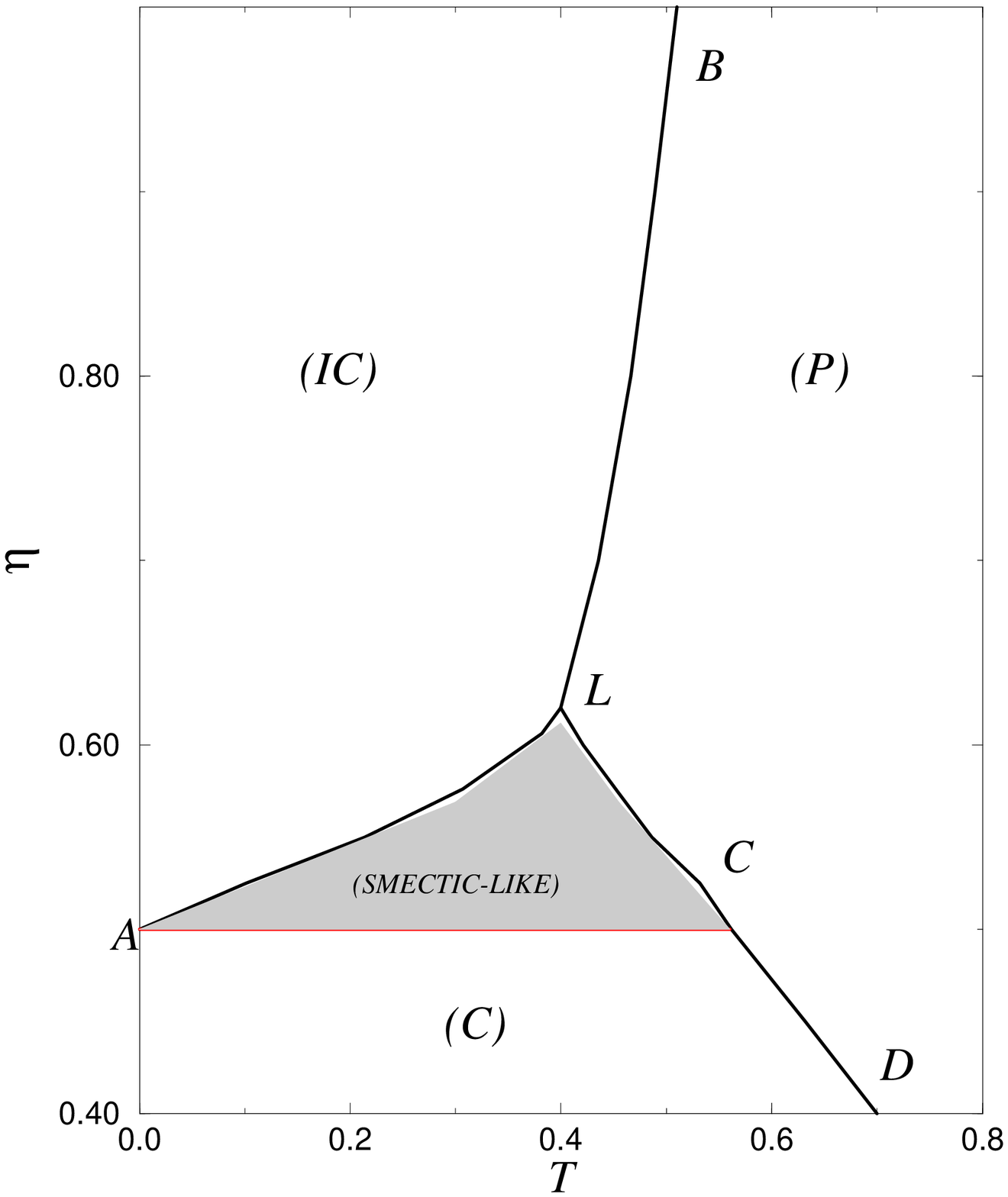}}
}
\end{center}

\protect\caption{
 MC phase diagram for the row model, in the ( $\eta, T$ ) plane.
}
\label{PHASEDIAG}
\end{figure}

\begin{figure}
\begin{center}
{\parbox[t]{11.5cm}{\epsfxsize 11.5cm
\epsffile{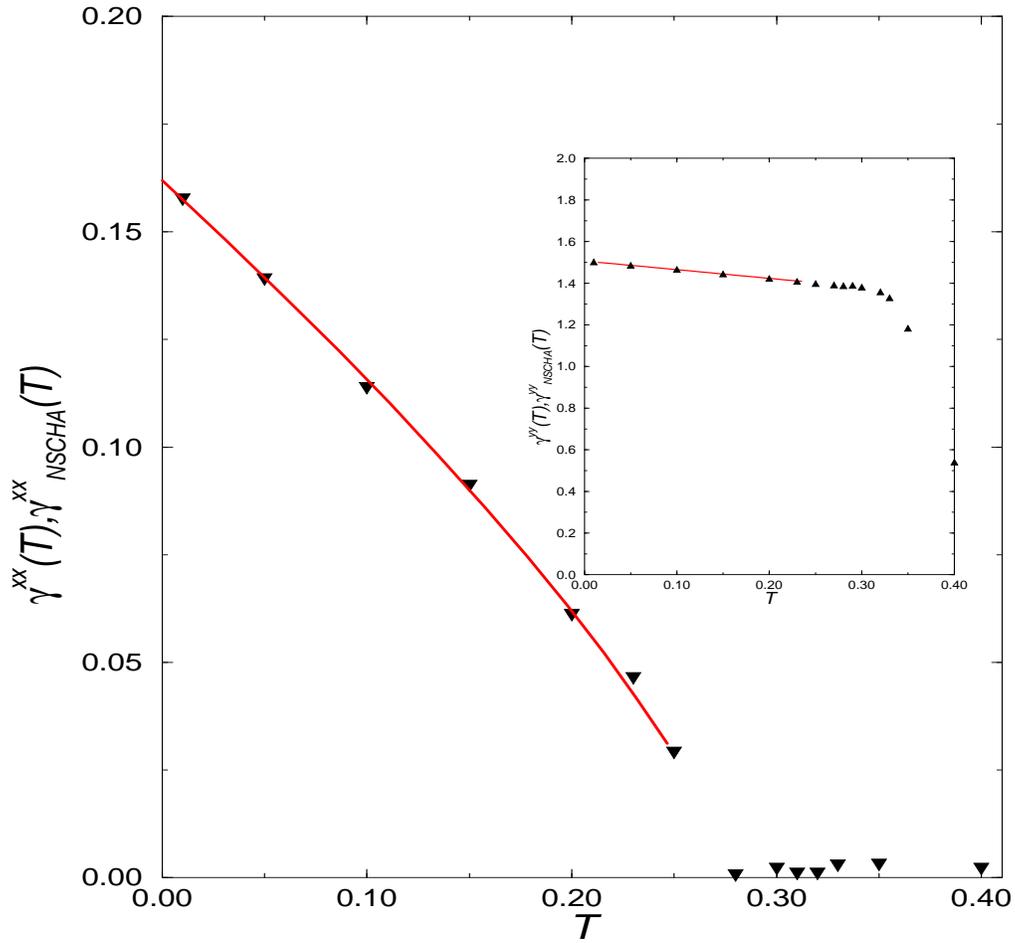}}
}
\end{center}
\protect\caption{
Monte Carlo stiffnesses in the $x$ and $y$ (inset) directions 
versus $T$ for the row model when $\eta=0.575$.
Triangles represent MC data, solid lines are the NSCHA predictions.
}
\label{NSCHA_TR}
\end{figure}

\begin{figure}
\begin{center}
{\parbox[t]{11.5cm}{\epsfxsize 11.5cm
\epsffile{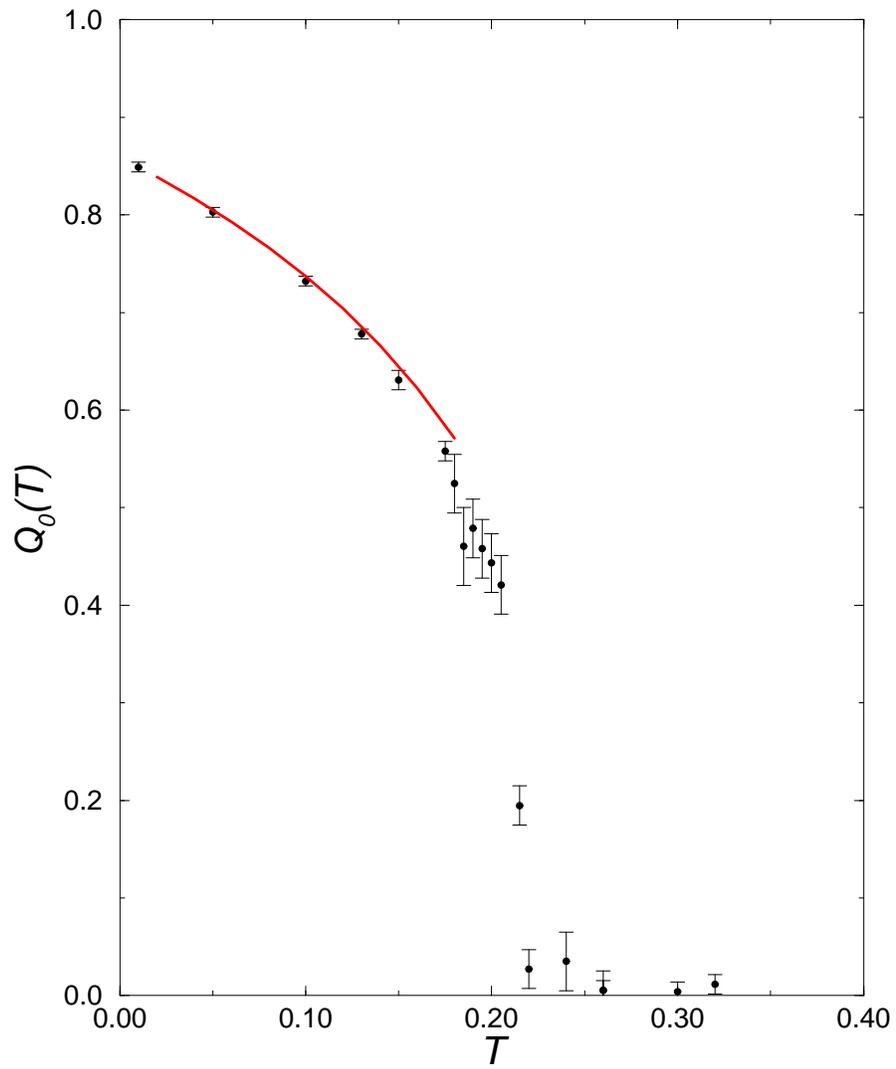}}
}
\end{center}
\protect\caption{
$Q_0(T)$ versus $T$ for $\eta=0.55$. Filled circles represent MC data and the 
solid line is the NSCHA prediction.
}
\label{Q0_055}
\end{figure}

 In the phase diagram of Fig (\ref{PHASEDIAG}), AL is a line separating the spiral
 incommensurate phase from the 
commensurate layered antiferromagnetic \textrm{(C)} phase. 
 It is characterized by a divergence of the chiral succeptibility and by the     \textit{%
continuous vanishing} of the $x$ component of the spin stiffness (Fig (\ref{NSCHA_TR})). The $y$
component of the spin stiffness, on the other hand, does not show any non-analiticity near AL.
In this part, we keep the value $\eta$ fixed and we vary the temperature.
 Typically we chose $\eta=0.575$ and
$\eta=0.55$. Starting from the low temperature phase, 
we observe that $\gamma^{xx}\to 0 $ and that simultaneously the chiral
 susceptibility diverges as one approaches AL, Fig (\ref{NSCHA_TR}).
 This behavior can be understood as follows:
Eikmans et al's Coulomb gas analysis of the generalized Villain model\cite{ETK},
 when generalized to the row model, gives\cite{FOOT}: 
\begin{equation}
 \gamma^{xx}\propto \frac 1{\chi _\sigma  }
\label{annula}
\end{equation}
Chiral variables and spin angle variables are coupled  in the (IC) phase; thus
$\gamma^{xx}$ can go to zero in a continuous fashion, rather than jump, on crossing
  AL. Similarly from the same Coulomb gas analysis,  one expects that $\gamma^{yy}$ is well
 behaved across AL (Fig (\ref{NSCHA_TR})).   
Fig (\ref{Q0_055}) shows that $Q_0$ also goes to zero (mod 2$\pi$) at $T_{C-IC}$.
 At first sight, 
the system appears to simply evolve from a homogeneous (IC) phase
into a homogeneous (C) phase as $T\to T_{C-IC}$ from below. 
If this picture were correct, here is what  histograms would yield: at low $T$,
$P(\vec \Delta)$ would display two maxima at $\pm \vec \Delta^0$
(corresponding to the two possible handedness of the spiral in the (IC) state).
As $T\to T_{C-IC}$, the two peaks would merge into a single peak, 
and, for
$T>T_{C-IC}$, $P(\vec \Delta)$ would be a gaussian, centered at $\Delta_x=0$.
By Eqs.\ref{gamma2} and \ref{freenerhist}, we would expect $\gamma^{xx}>0$ for
$T>T_{C-IC}$.
\vskip 0.5cm

\begin{figure}
\begin{center}
{\parbox[t]{13.5cm}{\epsfxsize 13.5cm
\epsffile{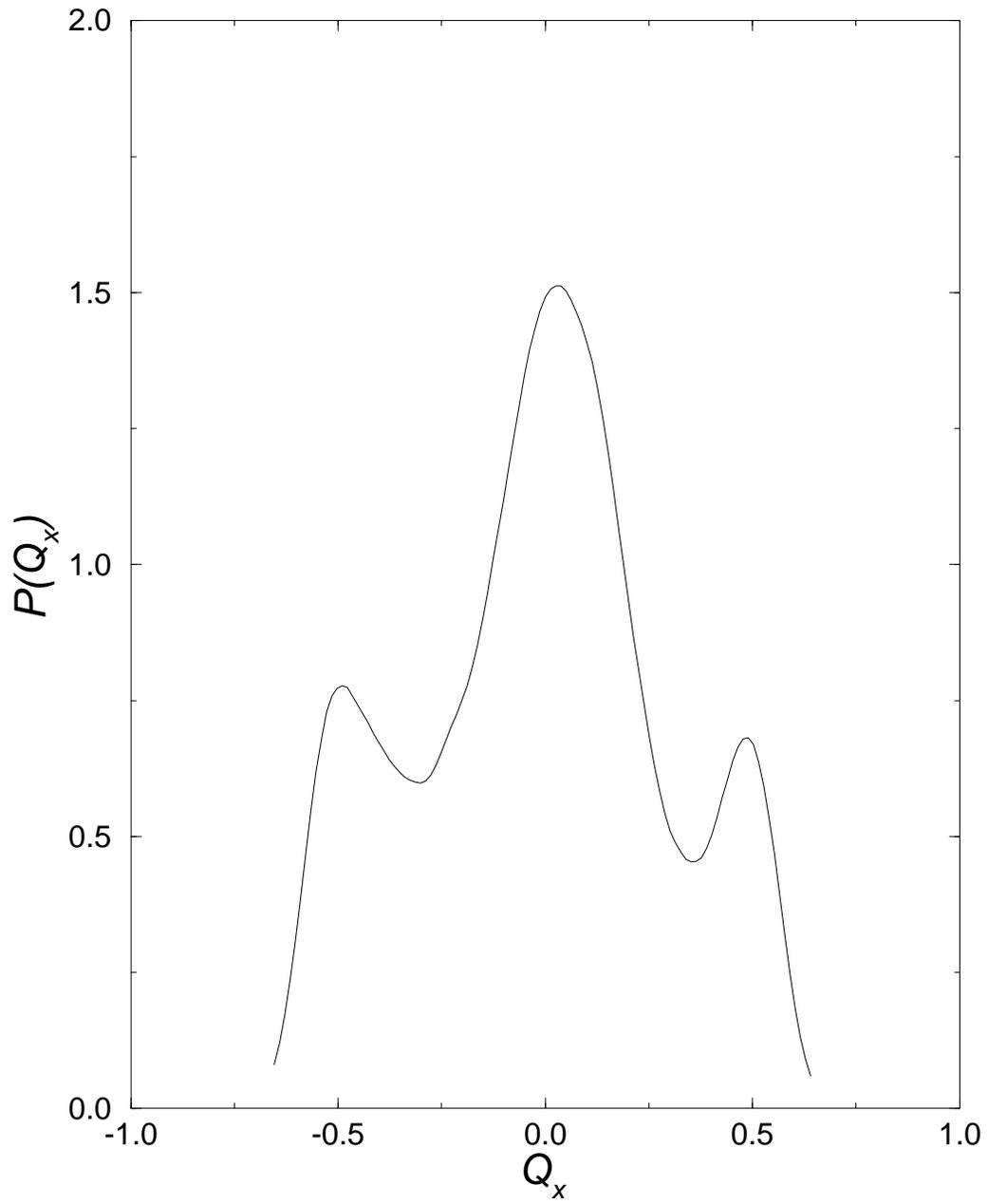}}
}
\end{center}
\protect\caption{
$P(Q_x)$ versus $Q_x$ for $\eta=0.55$ and $T=0.19J$.
}
\label{PQ}
\end{figure}

By contrast, here is what our simulation yields: at low $T$ we do get the two maxima
 at $\pm \vec \Delta^0$ and as 
$T\to T_{C-IC}$ they move closer to each other. However, they do not merge: the peaks at
$\pm \vec\Delta^0$ remain sharp and in addition a third peak develops at $\Delta_x=0$, such that
 $\Delta $-histograms show a three-peak structure for $T> T_{C-IC}$. There
 is a central peak at $\Delta_x=0$ and two side peaks centered at $T$
 dependent, finite values $\pm\Delta_0$. For sizes $48^2$ and for simulations
 using large enough MCS/spin the relative weight of the lateral
 peaks compared to the central peak is roughly one. Furthermore, this structure of the
$\Delta $-histogram is observed in a wide range of temperatures above $T_{C-IC}$. For instance we show
 the histogram for $T\gtrsim T_{C-IC}$  (Fig (\ref{PQ})).  
The structure of $P(\vec \Delta)$ could have two origins  :
 it could be associated with
a first order transition,  
and the fact that the multi-peak structure
survives for $T> T_{C-IC}$ could be linked to hysteresis effects, 
or it could be due to the occurence of a
non homogeneous thermodynamic phase. 
\vskip 0.5cm
\noindent 

The first order scenario is at variance with the observed temperature dependence of
 $\gamma _{xx}$ in two respects:

 {\it (i)} for $T\to T_{C-IC}$ from below, both $\gamma _{xx}$ and $\chi^{-1} _\sigma$ 
go {\it continuously} to zero,  as indicated by  Eq.\ref{annula}. {\it (ii)}
 For $T>T_{C-IC}$ up to the paramagnetic boundary, we find that
$\gamma_{xx}=0$ (see Fig (\ref{NSCHA_TR})); if we tried to explain this property in the
framework of a first order transition, this  would mean that the system is in a spinodal
state over a wide range of temperature,  
 which is rather unlikely.
\vskip 0.5cm
\noindent

Instead, we suggest that these data can be consistently interpreted if
 one considers the possibility of
a thermodynamically inhomogeneous phase for $T>T_{C-IC}$.
 We mentioned in 
the introduction that systems with competing interactions may lead to
 inhomogeneous groundstates consisting of ordered domains separated
 by domain walls\cite{GG,KASH,JH,UIP1,DW}. Our simulations reveal that
 the commensurate phase of the row model may well be such an example
 of stripe phases. 

From the shape of $P(Q_x)$ we see that a measure of
(the equilibrium value of) $Q_x$ at any point of the
 lattice gives $0$ with probability $1\over 2$, $+Q_0$ with probability $1\over
4$ and $-Q_0$ with probability $1\over 4$ ($+Q_0$ and $-Q_0$ are the secondary
maxima of $P(Q_x)$, see Fig (\ref{PQ})) The connection between the equilibrium
value of $Q_x$ and the plaquette chirality (Eq.\ref{sigmaprimeMC})
 implies that the chirality of any
site of a given sublattice will be positive, negative and zero with probability
$1\over 4$, $1\over 4$ and $1\over 2$ respectively:
indeed, if we had a homogeneous phase characterized by $Q_x=+Q_0$ over the entire system,
 the chiralites of the plaquettes of
 a given
 sublattice --denoted by $A$-- would all have the same sign, say positive;
 similarly, if we had $Q_x=-Q_0$ over
 the entire system, the chirality of $A$ would be negative for all the plaquettes;
lastly, if $Q_x=0$  over the entire system, the chirality of $A$ would be zero for all
the plaquettes. Since the values of $Q_x$ are distributed according to 
$P(Q_x)$, we deduce the above mentioned  distribution for the chiralities of any site of $A$. 

\begin{figure}
\begin{center}
{\parbox[t]{11.5cm}{\epsfxsize 11.5cm
\epsffile{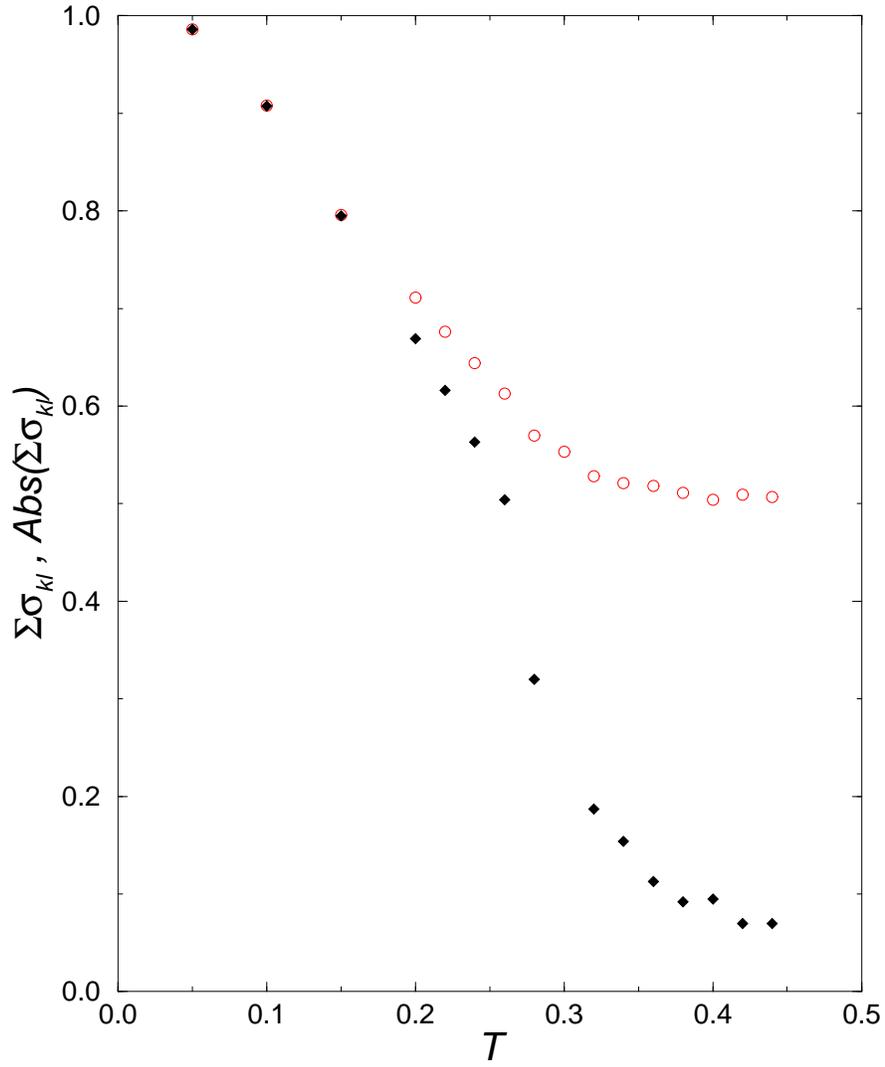}}
}
\end{center}
\protect\caption{
MC determination of the plaquette chirality  $\sum_{\left\langle kl \right\rangle
 \in P}\sigma_{kl}$ (filled diamonds) and of the absolute value of the plaquette
 chirality $Abs(\sum_{\left\langle kl \right\rangle \in P}\sigma_{kl} $ (open 
circles) versus $T$ for  $\eta=0.575$.
}
\label{SIG_ABSSIG}
\end{figure}

\noindent
Fig (\ref{SIG_ABSSIG}) precisely confirms this analysis. It is a 
plot (as a function of $T$) of the staggered chirality versus  $T$ (Eq.\ref{sigmaprimeMC}) and of the
 absolute value of the chiralities  (where we replace  
$\sum_{\left\langle kl\right\rangle \in P}\sigma _{kl}$ by
 $Abs\Bigl(\sum_{\left\langle kl\right\rangle \in P}\sigma _{kl}\Bigr)$
 in Eq.(\ref{sigmaMC})), for $\eta=0.575$. These two quantities give access to the number of plaquettes
 with positive, negative and zero chirality
one each sublattice (see Ref.\ref{LBZG}). 
We see that, for $T=0.4 J$,  well above
 the (C)-(IC) transition temperature, in  what should be the commensurate phase i.e a state with zero chirality,
 $25\%$ of the plaquettes have a positive chirality, $25\%$ of the plaquettes
 have a negative chirality, and $50\%$ of the plaquettes have no chirality.
 With these weights, averaging $Q_x$ over the system yields $Q_x=0$.

The shape of $P(Q_x)$ also signals the breakdown of the fluctuation-dissipation theorem. The correct
procedure required to extract the value of $\gamma^{xx}$ is to average 
$\frac{\delta ^2f(\vec \Delta )}{\delta \Delta
_x^2}$ over the distribution $P(\vec \Delta)$. If the dominant contribution to $P(\vec \Delta)$
comes from a single value, $\vec \Delta=\vec \Delta_0$, this yields Eq.\ref{gamma2}, which gives
the most probable value of  $\gamma^{xx}$. If $P(\vec \Delta)$ has a multi-peak structure,
 as is the case
here, Eq.\ref{gamma2} is not valid:
choosing for $\vec\Delta_0$ the value of $\vec \Delta$ corresponding to $Q_x=+Q_0$, or to $Q_x=-Q_0$,
or to $Q_x=0$ which is the mean value of $P(\vec \Delta)$, would give different values for 
$ \gamma^{xx}$ but such that $\gamma^{xx}>0$ .
 By contrast, the average of $\frac{\delta ^2f(\vec \Delta
 )}{\delta \Delta_x^2}$ over $P(\vec \Delta)$ leads to $\gamma _{xx}=0$ (see Fig (\ref{NSCHA_TR})).
So averages and most probable values do not coincide. 

The picture that emerges from the previous results is that of an inhomogeneous structure for
$T>T_{C-IC}$: domains of the spiral phase with pitch $+Q_0$ coexist with domains of 
the spiral phase with pitch $-Q_0$,
and the two types are separated by domain walls of the collinear phase. It is known that the
transition from a homogeneous phase (the (IC) state) to a domain structure can be
 continuous\cite{DN}, which is consistent with our results.   
The  spatial configuration of the domains is visualised in 
Fig (\ref{VISURUB}), which is a snapshot of the chiralities for
 $\eta=0.575$ and $T=0.4 J$. The morphology of the state is that of a striped phase.
 Note that the normal to the direction of the stripes correlates with $x$ 
 (the direction of the $\eta$-bonds). We dub this structure a smectic-like phase:
it is solid-like along $y$ ($\gamma^{yy}>0$) but has no rigidity
 along $x$ ($\gamma^{xx}=0$);
its effective free-energy in the hydrodynamic limit is similar to that of a smectic system
 (Ref(\ref{LDLK})).

\subsection{Study of the (C)-(IC) line at fixed $T$}
 To map out the domain of stability of
 the striped phase in the ($\eta,T$) plane, we keep $T$ fixed and we vary
 $\eta$. Fig (\ref{GAMAETA}) shows $\gamma _{xx}$ versus $\eta$ for $T=0.2 J$
 and $T=0.4 J$; in the region delimited by lines AC ($\eta=0.5$)
 and AL we get $\gamma _{xx}=0$ and one expects a striped phase there.
 In other words
AL separates an incommensurate phase from an inhomogeneous, non-collinear
  state.

 We have also sought for an analytical evidence of the inhomogeneous state 
in region ALC of the phase diagram Fig (\ref{PHASEDIAG}). The next section presents
results using NSCHA, a variational technique appropriate for frustrated systems: if
one seeks a uniform collinear solution in region ALC, one finds that $\gamma^{xx}<0$;
this behavior stems from the fact that the system is thermodynamically unstable with
respect to the formation of domains having either $Q_x=+Q_0$  or $Q_x=-Q_0$, the two
types connecting via domain walls of the collinear ($Q_x=0$) phase. The breakdown of
linear response and the properties of $\gamma^{xx}$
are hallmarks of the physics of dipolar magnets and of spin glasses\cite{GG,KASH,JH}. 

\section{NSCHA for the commensurate and\protect\\  incommensurate
regimes}\label{sec.NSCHA}

In a previous paper we introduced the new self-consistent harmonic
approximation (NSCHA)\cite{BZG}, a variational technique appropriate for frustrated
systems. The main feature of this approach is that it 
preserves the coupling between the chiral ground states of the system,
and that it takes long wavelength chiral fluctuations into account. Chiral and
phase (spin angle) variables remain coupled at all temperature T. We
now apply this method to
the row model.

\subsection{The NSCHA\ variational method.}

The Hamiltonian for XY\ spins characterized by spin angles $\left\{ \theta
_i\right\} $, reads 
\begin{equation}
H=-\sum_{\left\langle ij\right\rangle }J_{ij}\cos \left( \theta _i-\theta
_j\right)  \label{H1}
\end{equation}
where the $J_{ij}$ are nearest neighbor interactions. For frustrated systems
the sign of the product of the $J_{ij}$ over the links of a plaquette $P$
is negative and this may lead to non-collinear configurations in thermal
equilibrium. The variational method seeks to approximate $H$
(Eq.\ref{H1}) by an harmonic Hamiltonian $H_0$ . We rewrite the $\theta _i$
in Eq.\ref{H1} as  \begin{equation}
\theta _i=\theta _i^0+\varphi _i  \label{chg1}
\end{equation}
with $\theta _i^0=\left\langle \theta _i\right\rangle _{H_0}$ and 
\begin{equation}
H_0=\frac 12\sum_{\left\langle ij\right\rangle }\tilde J_{ij}\left( \varphi
_i-\varphi _j\right) ^2  \label{H0}
\end{equation}
Hamiltonian Eq.\ref{H1} is then mapped onto the NSCHA effective hamitonian
$H_{NSCHA}$\cite{BZG},
\begin{eqnarray}
H_{NSCHA} &=&-\sum_{\left\langle ij\right\rangle }J_{ij}\cos \left( \theta
_i^0-\theta _j^0\right) \cos \left( \varphi _i-\varphi _j\right) -\frac
1{2T}\sum_{\left\langle ij\right\rangle }\sum_{\left\langle kl\right\rangle
}J_{ij}J_{kl}  \label{HNSCHA} \\
&&\times \sin \left( \theta _i^0-\theta _j^0\right) \sin \left( \theta
_k^0-\theta _l^0\right) \sin \left( \varphi _i-\varphi _j\right) \sin \left(
\varphi _k-\varphi _l\right)   \nonumber
\end{eqnarray}

We then average Eq.\ref{HNSCHA} over $H_0$ (Eq.\ref{H0}) and minimize with respect to the variational
 parameters\cite{BZG} $\theta _i^0$ and $\tilde J_{ij}$ to obtain the NSCHA
variational equations.

In this ensemble we can compute the spinwave stiffness matrix. Its
eigenvalues are $\gamma _{NSCHA}^{xx}$ and $\gamma _{NSCHA}^{yy}$ : 
\begin{eqnarray}
\gamma _{NSCHA}^{xx} &=&\frac 1N\sum_{\left\langle ij\right\rangle
}J_{ij}\cos \left( \theta _i^0-\theta _j^0\right) (\vec u_{ij}.\vec
u_x)^2e^{-y_{ij}/2}  \label{gammaxx} \\
&&-\frac 1N\frac 1T\sum_{\left\langle ij\right\rangle }\sum_{\left\langle
kl\right\rangle }J_{ij}J_{kl}(\vec u_{ij}.\vec u_x)(\vec u_{kl}.\vec
u_x)e^{-(y_{ij}+y_{kl}+y_{ik}+y_{jl}-y_{il}-y_{jk})/2}  \nonumber \\
&&\times \left[ \cos \left( \theta _i^0-\theta _j^0\right) \cos \left(
\theta _k^0-\theta _l^0\right) +\sin \left( \theta _i^0-\theta _j^0\right)
\sin \left( \theta _k^0-\theta _l^0\right) \right]   \nonumber
\end{eqnarray}
where $\vec u_x$ is the unit vector in the horizontal direction, $\vec
u_{ij}$ is the vector connecting nearest neighbor sites $i$ and $j$
and
$y_{ij} =\left\langle (\varphi _i-\varphi _j)^2\right\rangle _{H_0}$
 For $%
\gamma _{NSCHA}^{yy}$ we replace $\vec u_x$ by $\vec u_y$ the unit vector in
the vertical direction.

\subsection{NSCHA\ for the row model}

Applying NSCHA to the row model gives two types of solutions

\noindent 
a) Commensurate solutions:\\ They are characterized by 
\begin{equation}
\theta _i^0-\theta _j^0={\vec Q^0}.\vec u_{ij}\;\left( \limfunc{mod}2\pi \right) 
\end{equation}
with 
\begin{equation}
Q^0_x=0;\;\;Q^0_y=\frac{2\pi }{\sqrt{3}}\;\;\left( \limfunc{mod}2\pi \right)
\end{equation}
and by nearest neighbor couplings $\tilde J_{ij}$. There are only two 
independent interactions namely $\tilde J_{ij}=\tilde \eta \tilde J$ for i
and j along the horizontal direction, and $\tilde J_{ij}=\tilde J$
otherwise. These satisfy the following equations
\begin{equation}
\tilde J=Je^{-\frac T{\pi \tilde J}\tan^{-1} \left[ \left( 1+2\tilde \eta
\right) ^{-1/2}\right] }  \label{Jtilde}
\end{equation}
\begin{equation}
\tilde \eta \tilde J=-\eta Je^{-\frac T{\pi \tilde \eta \tilde J}\tan^{-1}
\left[ \tilde \eta \left( 1+2\tilde \eta \right) ^{-1/2}\right] }
\label{etatilde}
\end{equation}
Eqs.\ref{Jtilde} and \ref{etatilde}\ can be self-consistently satisfied
without restriction for $\eta \leq 1/2$.

\noindent
 However, if $\eta >1/2$ 
equations Eqs.\ref{Jtilde} and \ref{etatilde}\ have no solution when $T\leq \frac J\eta \ln (2\eta )$;
this was to be expected, since the stable state of the system is a spiral structure at
low $T$, for $\eta >1/2$.

b) Incommensurate solutions:\\ They correspond to 
\begin{equation}
\theta _i^0-\theta _j^0={\vec Q^0}.\vec u_{ij}\;\;\left( \limfunc{mod}2\pi \right) 
\end{equation}
with 
\begin{equation}
Q^0_x=Q_0(T);\;\; Q^0_y=\frac{2\pi }{\sqrt{3}}\;\;\left( \limfunc{mod}2\pi \right)
\end{equation}

The variational equations can only be solved numerically. Just as for the FFTXY\ model, the
$\tilde J_{ij}$ are no longer short range interactions (for large $R$, 
 $\tilde J_{ij}\sim 1/\left|\vec r_i-\vec r_j\right| ^6$
 see Ref(\ref{LBZG})) and 
the sign of $\tilde J_{ij}$ varies with the relative orientation of $i$
and $j$. 

Knowledge of the $\tilde J_{ij}$ allows us\cite{BZG} to compute the free-energy, $Q_0(T)$ , $\gamma
_{NSCHA}^{xx}$ , $\gamma _{NSCHA}^{yy}$ (Eq.\ref{gammaxx}) and the staggered chirality $\sigma
_{NSCHA}$ as a function of $T$ {\it for all $\eta$}. 

For {\it all} $\eta<0.5$ the lowest free-energy is obtained for the commensurate solution and up to
the (C)-(P) boundary (line CD in Fig (\ref{PHASEDIAG})) $\gamma_{NSCHA}^{xx}>0$.

For $0.5<\eta<\eta_L$, the (IC) solution has the lowest free-energy at low $T$ ($T<T_{C-IC}$).
 As seen on Figs (\ref{NSCHA_TR})
and (\ref{Q0_055}), 
NSCHA and MC results agree closely except in the vicinity of ${T}_{C-IC}$, where
defects are expected to play an important role (see our previous paper Ref.\ref{LBZG}). In that
regime $\gamma_{NSCHA}^{xx}>0$. For $T>T_{C-IC}$ the variational equations favor a commensurate
configuration, but we find  that $\gamma_{NSCHA}^{xx}<0$: the solution is thermodynamically 
unstable. By this we mean that NSCHA yields a (C) solution in region ALC of the phase diagram, but that
fluctuations around the solution (given by $\gamma_{NSCHA}^{xx}$)  generate an instability.

\vskip 1.5cm
To summarize our results,
\begin{enumerate}
\item in the $\eta,T$ plane, the transition between the spiral phase  and the (C) phase is only seen
 at point A (that is, at zero temperature). Consequently, the Lifshitz point
 is at $T=0$ for the 2D XY model. 

\item the existence of the striped phase suggests that chiral variables and phase variables
 remain strongly coupled at all $T$. This may explain why, despite the fact that
 $\gamma _{xx}=0$ in the striped phase, vortices do not unbind (leading to a reentrant
 paramagnetic phase).
The relevance of this coupling had already been emphasized in our study of the fully frustrated case
 ($\eta=1$). 

\item 
the existence of the inhomogeneous state affects scaling analyses near the
 (IC)-stripe phase boundary (line AL): Fig (\ref{BINDEROP}) shows the Binder order parameter
 Eq.(\ref{binder}) as a function of $T$ for $\eta=0.575$. We do not observe a clear intersection
 at the critical temperature. A similar feature had been pointed out by Olsson in his study
 of fully frustrated XY spins on a 2D square lattice (Ref.\ref{LO3}).

\item because the striped phase is spatially inhomogeneous, it is not easy to 
define appropriate boundary conditions for the MC simulation. Uniform twists will produce frustration.

\end{enumerate}
The present work has revealed the existence of a smectic-like phase. This raises the question
 of the nature of the transition between the striped phase and the (P) phase (line LC)
 and also between the striped phase and the commensurate phase  (line AC); for instance,
 if the transition line LC is not KT-like, one also needs to understand the nature of
 the critical regime along CD: for $\eta<<0.5$ one recovers a KT-transition so there has to
 be some cross-over. Work is in progress to clarify that issue.

\section{Conclusion}
Using detailed Monte Carlo simulations 
we have studied the
commensurate-incommensurate transition
of the two dimensional XY model on a triangular lattice. Our study shows that
this transition only occurs at $T=0$. At finite temperature, the incommensurate
structure evolves into a striped phase made up  of domains of left- and 
right-handed spirals separated by walls. The domain walls consist of
the collinear structure. This state resembles the smectic-A phase of
liquid crystals. The nature of the phase transitions between the 
striped phase and the ordered phases or between the striped phase and the
paramagnetic phase is an open problem. Analytical calculations using
NSCHA ( a variational approach well suited for non collinear structures
) support the MC results. The connection between
frustrated XY models and the vortex state of layered type II
superconductors suggest to view the smectic phase as a vortex liquid
state. This regime would appear to be an intermediate phase between the superconducting
and the metallic states, critical in one subspace and quasi-ordered in the
other.

\vspace{2.0cm}

\section*{ACKNOWLEDGMENTS}

Monte Carlo calculations were performed on a Cray C98 thanks to contrat
960162 from IDRIS. Support from NATO grant 930988 is acknowledged.

\newpage
\vspace{-2truecm}
\newpage
\newpage

\newpage
\vskip -2.0cm
\newpage


\begin{thebibliography}{9}
\bibitem{FOOT1} There is a vast literature on this problem;  see for instance
references quoted in the following four articles:
\bibitem{FOOT} Eq.\ref{annula} can be understood with the help of the following
argument: using definition (\ref{sigmaprimeMC}),  the staggered chirality order parameter
 $\Sigma$ (Eq.\ref{sigmaMC}) is expressed as a sum over nearest neighbor phase differences
 in the $x$ direction. This sum is proportional to $Q_x$, so that $\chi _{\sigma}$
 (Eq.\ref{chiralxi}) contains the fluctuations of $Q_x$ described by $\chi _{Q_x}$.
 By Eq.\ref{gammadelta} one gets Eq.\ref{annula}
\end{thebibliography}
\end{document}